\documentclass[aps, pra, twocolumn, showpacs,footinbib,superscriptaddress]{revtex4-1}
\usepackage{graphicx, subfigure,braket}
\usepackage{amsmath,amssymb}
\usepackage{color}
\usepackage{natbib,hyperref}
\usepackage{placeins}

\begin{document}

\title{Two-qubit {\sc cz} gates robust against charge noise in silicon while compensating for crosstalk using neural network}

\author{David W. Kanaar}
\affiliation{Department of Physics, University of Maryland Baltimore County, Baltimore, MD 21250, USA}
\author{Utkan G\"ung\"ord\"u}
\affiliation{Laboratory for Physical Sciences, College Park, Maryland 20740, USA}
\affiliation{Department of Physics, University of Maryland, College Park, Maryland 20742, USA}
\author{J.~P.~Kestner}
\affiliation{Department of Physics, University of Maryland Baltimore County, Baltimore, MD 21250, USA}

\begin{abstract}

The fidelity of two-qubit gates using silicon spin qubits is limited by charge noise. When attempting to dynamically compensate for charge noise using single-qubit echo pulses, crosstalk can cause complications. We present a method of using a deep neural network to optimize the components of an analytically designed composite pulse sequence, resulting in a two-qubit gate robust against charge noise errors while also taking crosstalk into account. We analyze two experimentally motivated scenarios. For a scenario with strong EDSR driving and negligible crosstalk, the composite pulse sequence yields up to an order of magnitude improvement over a simple cosine pulse. In a scenario with moderate ESR driving and appreciable crosstalk such that simple analytical control fields are not effective, optimization using the neural network approach allows one to maintain order-of-magnitude improvement despite the crosstalk.
\end{abstract}

\maketitle

\section{Introduction}
Silicon devices, such as those using SiMOS and Si/SiGe, are a promising platform for scalable quantum computation due to the small size of the qubits and existing industrial infrastructure\cite{Zwanenburg2013}. Arbitrary quantum computation requires the ability to perform a universal set of quantum gates  with low error\cite{Divincenzo2000}. Although error correcting codes with concatenation can correct long calculations to arbitrary accuracy they still require very small initial error rates per gate, on the order of $10^{-4}$ \cite{NielsenandChuang}, even in the case of surface codes if one wishes to avoid a huge overhead in the number of physical qubits \cite{Fowler2012}. In silicon, single-qubit gates with infidelities of $10^{-3}$ have been achieved \cite{Yang2019,Yoneda2018,Petit_2020}, however a universal gate set also must contain an entangling gate. State-of-the-art two-qubit gate fidelities \cite{xue2021,noiri2021fast,mills2021twoqubit,Huang2019} have not reached the same level, mainly because of charge noise and crosstalk. Crosstalk in the context of this paper will refer only to an insufficiently large separation between the resonant frequencies of two qubits (compared to the strongest driving used, i.e., the maximum Rabi frequency $\Omega_{\text{max}}$), leading to unwanted rotation of an idle qubit while another qubit is resonantly driven. Other effects are sometimes also referred to as crosstalk, such as heating, rectification effects changing qubit frequency, and capacitive interaction changing qubit parameters, are not taken into account here. This is because the mechanisms behind most of these effects are not characterized well enough for us to accurately model them in our Hamiltonian. 

In this paper we theoretically design a high-fidelity two-qubit controlled-Z ({\sc cz}) gate in silicon that is robust against charge noise induced exchange fluctuations and Stark shifts in the presence of crosstalk. Although silicon is seen as a scalable platform for quantum computing because of the small physical size of the qubits, crosstalk becomes a challenge when multiple qubits are present. The simplest way to circumvent crosstalk is to create a large difference, $\Delta E_z$, between the Zeeman energy of the target qubit and that of the idle qubits, but that is challenging to scale up \cite{seedhouse2021}. Methods of dealing with crosstalk without dynamically correcting exchange fluctuations have been recently proposed \cite{Hansen2021,Heinz2021}. Other work has considered crosstalk while dynamically correcting exchange fluctuations \cite{Gungordu2020}, but absent fluctuations in Zeeman energy. On the other hand, a method of dynamically correcting fluctuations in both exchange and Zeeman energies in two-qubit gates has been proposed \cite{utkan2018} assuming crosstalk is negligible. This paper fills the gap in the literature by simultaneously compensating crosstalk, exchange fluctuations, and Zeeman energy fluctuations using a combination of a composite pulse sequence and neural network shaped segments.

\section{Model}
The two qubit system consists of two exchange coupled quantum dots in silicon. The system is known to be well described by an extended Heisenberg model \cite{Loss1998}:
\begin{multline}
    H_0= \frac{J(V)}{4} (XX+YY+ZZ-II)\\
    +\mu_B g_1 (B_{x,1} XI +B_{z,1} ZI)\\
    +\mu_B g_2 (B_{x,2} IX +B_{z,2} IZ   ),
    \label{eq:H}
\end{multline}
where $J(V)$ is the exchange coupling strength as a function of the barrier voltage $V$, $X,Y,Z$ are the Pauli matrices with Kronecker product assumed, $\mu_B$ is the Bohr magneton, $g_i$ is the electron $g$-factor of the $i$-th qubit and $B_{a,i}$ is the magnetic field in the $a$-direction at the $i$-th qubit. Changes in the barrier gate voltage change the position of the dots, and in experiments with position dependent magnetic fields this would change the fields at the dots if other plunger gates are not used to compensate to keep the dots in place as in Ref.~\cite{xue2021}. Additionally the barrier gate voltage is observed to affect the resonant frequencies of the qubits \cite{xue2021,UNSWprivate} via spin-orbit interaction \cite{TanttuSpin}. To take this into account, device-dependent functions $f_1(V)$ and $f_2(V)$ are introduced to the Hamiltonian, $H_1=H_0 + f_1(V) ZI+ f_2(V) IZ$. In experiments, the Zeeman energies, defined as $E_{z,i}=\mu_B g_i B_{z,i}$, are usually large, on the order of $h \times 10$GHz where $h$ is Planck's constant, compared to the other terms, on the order of $ h \times 10$MHz or less, in the Hamiltonian \cite{xue2021}. To simplify the Hamiltonian, one typically moves to the rotating frame defined by $R=e^{\frac{i}{2} (E_{z,1} ZI+E_{z,2} IZ) t}$. After the rotating wave approximation (RWA), the rotating frame Hamiltonian,  $H_R=R H R^{\dagger}+i \hbar (\partial_t R) R^{\dagger}$, becomes

\begin{multline}
    H_R \simeq \frac{J(V)}{4} ZZ+ f_1(V) ZI+ f_2(V) IZ\\
    + \frac{\Omega_1}{2} \left(\cos\phi_1 XI
    +\sin\phi_1 YI \right) + \frac{\Omega_2}{2}
    \left(\cos\phi_2 IX
    +\sin\phi_2 IY\right)
    \label{eq:HR}
\end{multline}
where the oscillating magnetic field along $x$ seen by the electron in the lab frame is applied by electron spin resonance (ESR) or electron dipole spin resonance (EDSR) and is taken to be composed of two tones at the two resonant frequencies, $B_{x,i}=\Omega_i \cos( E_{z,i} t+\phi_i)$. The approximation in the rotating frame of neglecting the fast counter-rotating terms has already been applied in the above.

Using the Hamiltonian in Eq.~\eqref{eq:HR} it is quite simple to create a naive {\sc cz} gate, equivalent to $e^{i \frac{\pi}{4}ZZ}$, by simply turning off the ESR/EDSR field and pulsing $J(V)$ adiabatically (so as not to violate the RWA) such that the area of the exchange pulse is equal to $\pi/2$. This type of {\sc cz} gate is distinct from a CROT gate\cite{Zajac_2018} which requires single-qubit driving at exchange-shifted resonances. In an adiabatic {\sc cz} gate, any residual $IZ$ and $ZI$ rotations from $f_1(V)$ or $f_2(V)$ can then be compensated with virtual $z$-rotations. However Eq.~\eqref{eq:HR} does not take charge noise or crosstalk into account and therefore such a naive {\sc cz} gate will not perform well if either is significant. We will use more sophisticated methods of designing the gate, presented in Sec.~\ref{sec:methods} in the interest of a self-contained discussion.

Charge noise causes fluctuations in the electrostatic environment of each dot, equivalent in their effect to fictional fluctuations in the gate voltages \cite{Reed_2016}, which in turn causes fluctuations in the exchange strength $J(V)$ as well as fluctuations in the $g$-factors. The fluctuations in exchange are reflected in Eq.~\eqref{eq:H} by taking $J(V) \longrightarrow J (V)+\frac{dJ}{dV}\delta V$. The $g$-factors depend linearly on the electric potential with $\Delta g \lesssim 0.002/V$ \cite{Ruskov2018} and effective potential fluctuations are on the order of $10\mu V$ \cite{HuangP2014,Orus2019,utkan2018}, resulting in fluctuations in the $g$-factors of $\Delta g \sim 10^{-8}$. These fluctuations of $g$-factor do not result in large errors for typical ESR or EDSR strengths less than $h \times 10$MHz. However the Zeeman energies can be on the order of $h \times 10$GHz so even small $g$-factor fluctuations are amplified by the large uniform magnetic field and need to be taken into account by changing $f_i(V)  \longrightarrow f_i (V+\delta V)$. Additionally, isotopic impurities in the silicon can lead to nuclear spin noise resulting in unwanted Overhauser fields which can be taken into account by adding $\delta E_{z,1} ZI +\delta E_{z,2} IZ $ to the Hamiltonian. So, there are noise terms in the Hamiltonian on the generators $IZ$, $ZI$, and $ZZ$. This noise will be dealt with in Sec.~\ref{subsec:EDSR} assuming the RWA of Eq.~\eqref{eq:HR} holds.

However, if crosstalk is significant, meaning $\Delta E_Z \equiv E_{z,1}- E_{z,2}$ is on the order of $J$ or $\Omega_i$, the RWA is not a good approximation. In that case one must work with a significantly more complicated Hamiltonian, as considered in detail in Sec.~\ref{subsec:ESR}. 

\section{Methods}\label{sec:methods}
\subsection{Pulse sequence}\label{subsec:sequence}
To correct for fluctuations in exchange it is possible to use a pulse sequence as shown in Ref.~\cite{utkan2018},
\begin{equation}\label{eq:seq}
     U_{\text{seq}}=e^{-i \eta IX }e^{-i \zeta ZZ}   e^{i \frac{\theta}{2} IX }  e^{-i \frac{\pi}{2}   ZZ} e^{-i \frac{\theta}{2} IX }  e^{-i \zeta ZZ} e^{i \eta IX }, 
\end{equation}
where the angles are defined through the relations $\zeta=-\frac{\pi}{4} \sec \theta$,  $\sec \theta = \frac{2}{\pi} \text{sinc}^{-1} \frac{\sqrt{2}}{\pi}$ and $\tan \eta = \tan \theta \sec \left( \frac{\pi}{2} \sec \theta \right)$. In Eq.~\eqref{eq:seq} the single-qubit driving was chosen to be on the second qubit, however, the first qubit would also have been a valid choice.\par
This composite sequence contains segments where the exchange is turned on and off adiabatically while the ESR/EDSR remains off, resulting in two-qubit entangling $ZZ$ phases, and segments where the exchange remains off while the ESR/EDSR is turned on and single-qubit $x$-rotations are performed.

Additionally, errors caused by fluctuations in Zeeman energy during the $ZZ$ segments can be canceled by splitting those segments in half and inserting $\pi$-pulses about $x$ on both qubits to echo out the Zeeman error,
\begin{multline}
     U_{\text{seq},\text{echo}}=e^{-i \eta IX }\cdot e^{-i \frac{\zeta}{2} ZZ}\cdot XX \cdot e^{-i \frac{\zeta}{2} ZZ}\cdot XX \cdot e^{i \frac{\theta}{2} IX } \cdot e^{-i \frac{\pi}{4}    ZZ}\\
     \cdot XX\cdot e^{-i \frac{\pi}{4}    ZZ}\cdot     XX \cdot e^{-i \frac{\theta}{2} IX }\cdot e^{-i \frac{\zeta}{2} ZZ}\cdot XX \cdot e^{-i \frac{\zeta}{2} ZZ}\cdot XX \cdot e^{i \eta IX } 
     \label{eq:seqfull}
\end{multline}

During each entangling $ZZ$ segment, $J(V)$ is to be pulsed such that the evolution is described by the adiabatic Hamiltonian, 
\begin{equation}
H_{ad} = 2 \pi \left[J(V) \frac{ZZ}{4}+ \sqrt{(J(V))^2+\Delta E_{z}^2} \frac{ZI-IZ}{4}  \right],
\label{eq:Had}
\end{equation}
where $\Delta E_z=E_{z,1}-E_{z,2}$ is the difference in Zeeman energies between the two qubits. By pulsing $J(V)$ adiabatically the small fluctuations in $J(V)$ caused by charge noise cause the resulting error channels to be limited to $ZI$, $IZ$ and $ZZ$. Likewise, fluctuations in Zeeman energies also result in $ZI$ and $IZ$ errors.

This pulse sequence does, however, assume that the single-qubit $x$-rotations are error free, which is not physical. Nevertheless, if the single-qubit $x$-rotations are engineered to be robust against charge noise, the total pulse sequence will be robust against charge noise. To find good robust single-qubit rotations and adiabatic pulses to compose the pulse sequence, a neural network approach will be used, as explained in the next section.

\subsection{Neural Network}\label{subsec:nn}
To find a robust pulse shape while also canceling crosstalk for two qubits requires that we include non-RWA corrections to Eq.~\eqref{eq:HR}. Handling the resulting Hamiltonian (discussed in Sec.~\ref{subsec:ESR}) analytically is difficult because the terms in the Hamiltonian do not commute in a way such that they form a simpler subalgebra of $\mathfrak{su}(4)$ such as $\mathfrak{su}(2)$. Therefore the error resulting from this Hamiltonian appears on the entirety of $\mathfrak{su}(4)$ which is 15-dimensional. Consequently, canceling errors in such a large space is handled numerically through optimization of a neural network instead of analytically.

The cost function to be minimized is a weighted sum of two terms. The first term, which makes sure the desired gate is performed in the absence of noise, up to single-qubit $z$-rotations which can be performed virtually \cite{Mckay2017}, is the noiseless infidelity $1-F$:
\begin{equation}\label{eq:costtarget}
    \min_{\vec{\varphi}} \left[1-\frac{1}{4}\left|\text{Tr}\left(e^{-i( \frac{\varphi_1}{2}ZI + \frac{\varphi_2}{2}IZ)} U_c e^{-i (\frac{\varphi_3}{2}ZI+ \frac{\varphi_4}{2}IZ)} U_{\text{t}}^{\dagger}\right)\right|^2\right]
\end{equation} 
where $U_{\text{t}}$ is the targeted rotation of the optimization and $U_c$ is the noiseless evolution operator calculated numerically using the control Hamiltonian of that optimization problem. The adiabatic infidelity is the same noiseless infidelity where $U_{t}$ is replaced with the adiabatic evolution operator, $U_{ad}$, calculated with the same control field as $U_c$.

The second term of the cost function is the Frobenius norm of first order Magnus term of the noise propagator, $\mathcal{E}$, and it determines how robust the pulse is against the effect of noise. $\mathcal{E}$ is calculated by treating the error perturbatively such that the Hamiltonian is described by $H=H_c+\epsilon H_{\epsilon}$ where all error terms are in $H_{\epsilon}$. For small errors $\epsilon$ the full evolution operator then becomes
\begin{equation}
    U \approx U_c e^{i \epsilon \int U_c(t) H_{\epsilon}(t) U_c^{\dagger}(t)dt}=U_c e^{i \epsilon \mathcal{E}},
\end{equation}
where $\mathcal{E}=\int U_c(t) H_{\epsilon}(t) U_c^{\dagger}(t)dt$.

\begin{figure}
  \centering
  \includegraphics[width=.9\columnwidth]{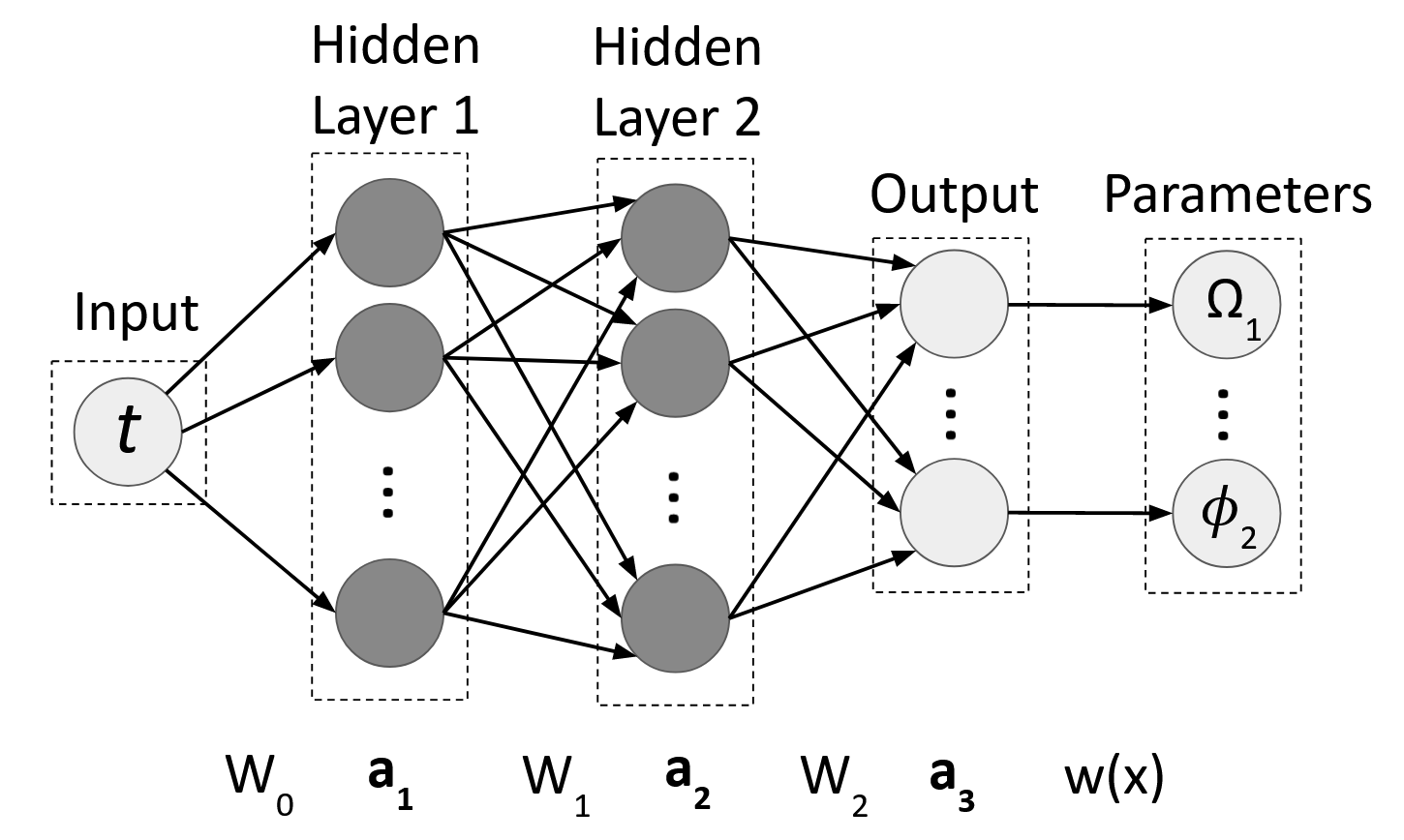}
  \caption{Schematic representation of a neural network where the input, output and hidden layers are shown as circles and the weights are shown as arrows.}\label{fig:NNschematic}
\end{figure}

We follow Ref.~\cite{Gungordu2020p2} in using a deep neural network to create our pulse shapes for robust and crosstalk resistant gates. A neural network can have many free parameters, in our case up to 6000, while still allowing rapid optimization due to the fact that the smooth shapes allow the use of efficient adaptive ODE solvers and the fact that the the gradient of the cost function is efficiently evaluated using automatic differentiation via back-propagation. The latter results from the structure of a neural network, which is a function that takes an input vector, in our case time $t$, and gives an output vector calculated through hidden layers, vectors $\mathbf{a}_i$, defined as
\begin{equation}
    \mathbf{a}_{i+1}= v(W_{i}\mathbf{a}_{i}+\mathbf{b}_i),
    \label{eq:NN1}
\end{equation}
where $W_{i}$ is the weight matrix between the hidden layers $\mathbf{a}_i$ and $\mathbf{a}_{+1}$, $\mathbf{b}_i$ is the vector of biases and $v(x)$ is a nonlinear activation function which acts on each element. Figure \ref{fig:NNschematic} shows a representation of the hidden layers as circles and the connecting weights as arrows.
In our case, a single element input, time $t$, is used and up to 5 elements of output were calculated through two hidden layers of size 32 with an activation function $v(x)=\tanh(x)$. The final output layer has no activation function and is therefore not limited in magnitude or sign. Some of the corresponding physical outputs of the neural network are however actually limited. The exchange $J(V)$ is always positive and bounded by the device's operational range therefore the output for $J(V)$, $a_{3,J}$, is fed into a wrapping function such that the final output is bounded, $J(V) = w_J(a_{3,J})=J_\text{max} \sin^2(a_{3,J})$. Additionally the ESR or EDSR driving amplitudes are also limited to the device's maximum but since they are allowed to be negative they are wrapped with $w_{\Omega}(x)=\Omega_\text{max} \sin(x)$. The ESR/EDSR driving phases $\phi_i$ are not physically bounded and so do not need to be wrapped.
The neural network was implemented in the Julia programming language using the DiffEqFlux.jl package \cite{rackauckas2019diffeqfluxjl}. The BS5 solver from the the OrdinaryDiffEq.jl package \cite{Rackauckas2017} was used to solve the Schrodinger equation and find the value of $\mathcal{E}$. Finally, the RADAM and BFGS optimizers from the Flux.jl and Optim.jl packages were used for optimizing the neural network.  

\section{Results}\label{sec:results}
\subsection{EDSR device with negligible crosstalk}\label{subsec:EDSR}
In this section we consider the case of a device with a large Zeeman energy splitting, $\Delta E_z=h \times 103$MHz \cite{xue2021}, and choose values of $\Omega_{\text{max}}=h \times 8$MHz and $J_\text{max}=h \times 10$MHz such that the RWA is valid and Eq.~\ref{eq:HR} can be used without limiting the fidelity. For this device the exchange depends exponentially on the barrier voltage $J(V)=J_0 e^{2 \alpha V}$ with  $J_0=h \times 0.058$MHz and $\alpha=12.1/$V. The barrier-induced Stark shifts of the resonant frequencies are $f_i(V)=V^\gamma  \beta_i $, with $\gamma=1.2$, $\beta_1=h \times -2.91$MHz$/V^\gamma$ and $\beta_2=h \times 67.1$MHz$/V^\gamma$ \cite{xue2021}.

\begin{figure}
  \centering
  \includegraphics[width=.9\columnwidth]{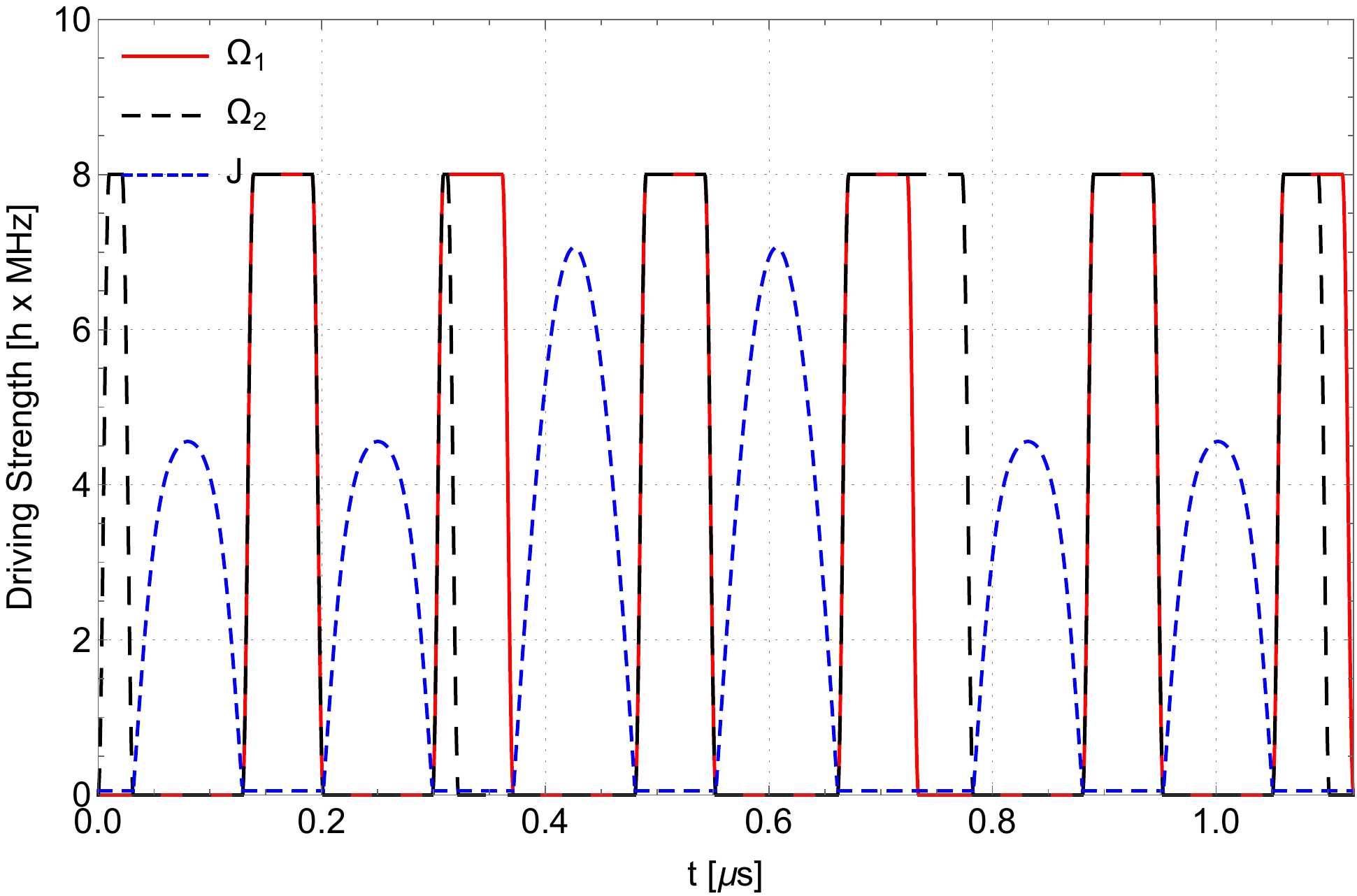}
  \includegraphics[width=.9\columnwidth]{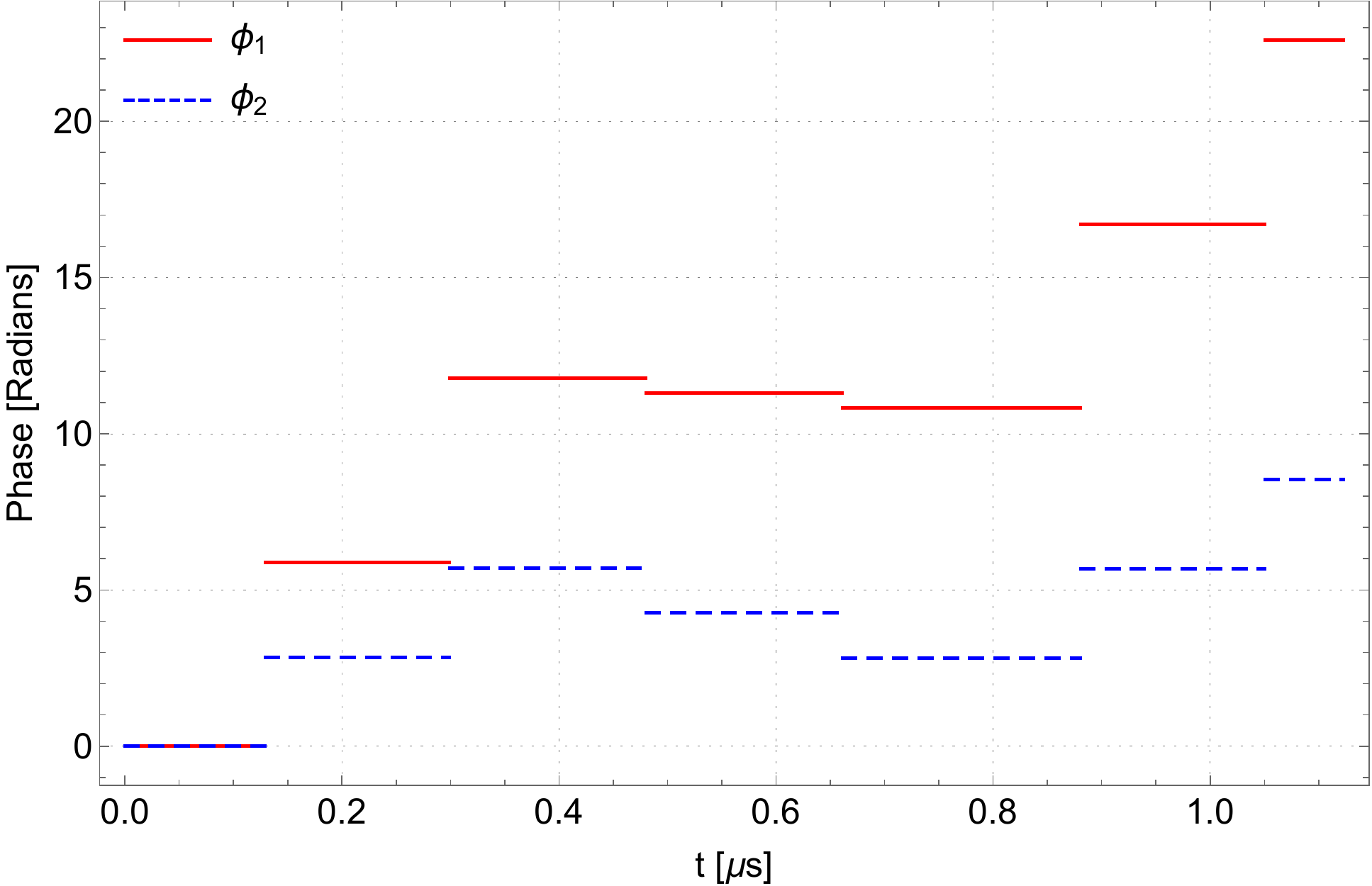}
  \caption{Analytical control fields vs time for the pulse sequence of Eq.~\eqref{eq:seqfull} when crosstalk is negligible. Top: Exchange $J$, and EDSR tone amplitudes $\Omega_1$ and $\Omega_2$ where $h$ is Planck's constant. Bottom: Accompanying phases of the EDSR tones.}\label{fig:VDSpulseshape}
\end{figure}

\begin{figure}
  \centering
  \includegraphics[width=.9\columnwidth]{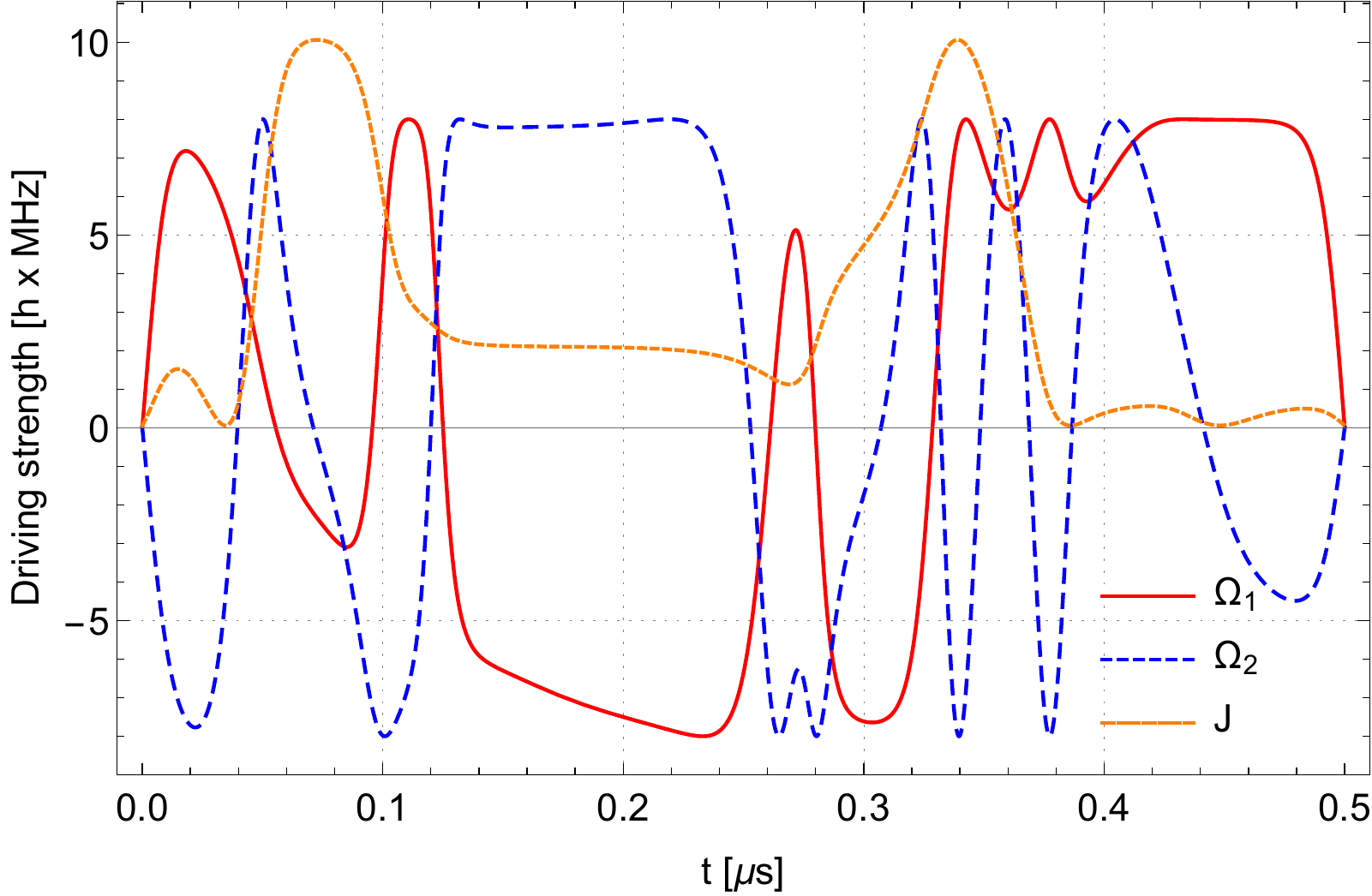}
  \includegraphics[width=.9\columnwidth]{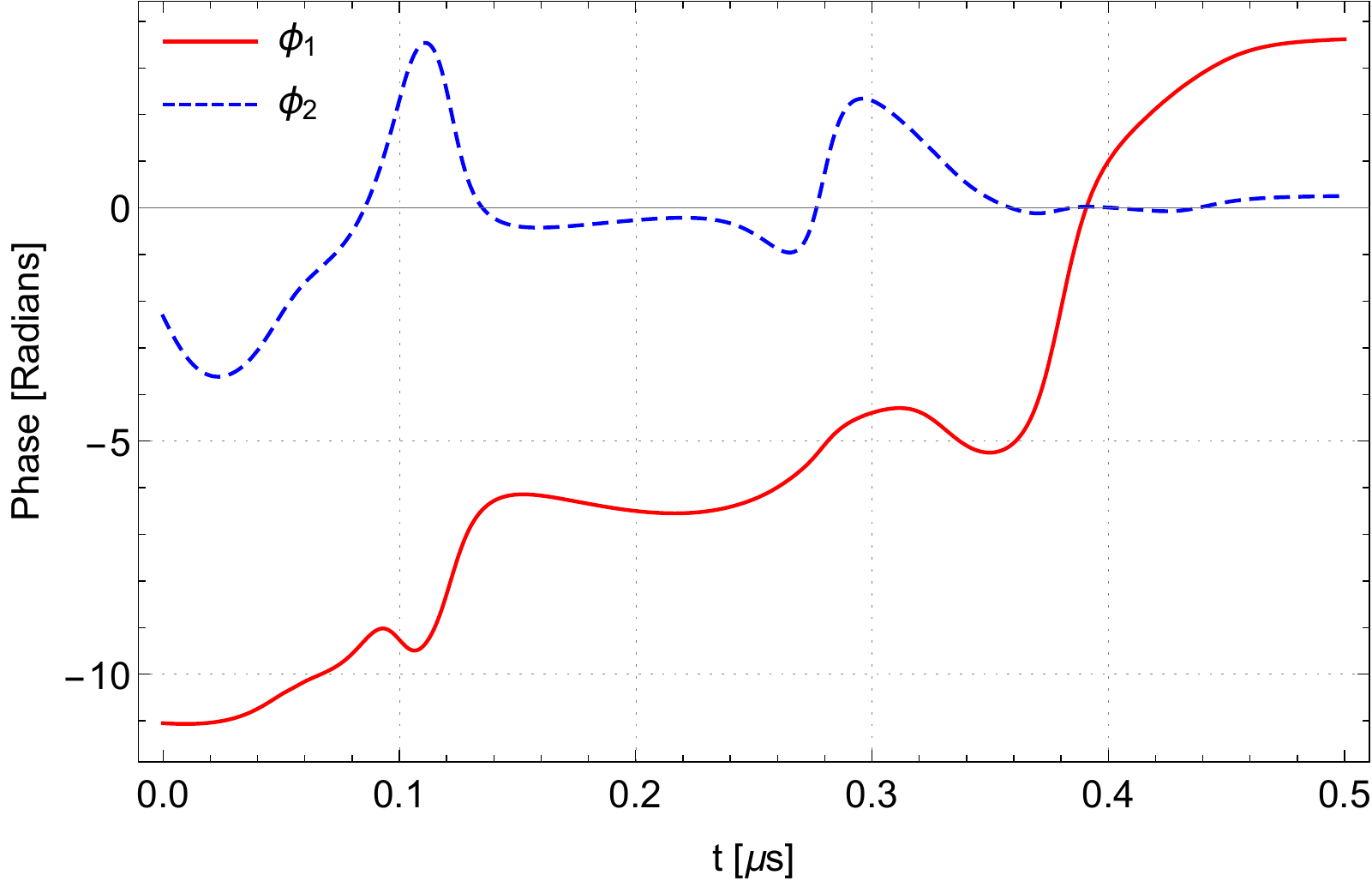}
  \caption{Neural network designed control fields vs time for
single shot pulse when crosstalk is negligible. Top: Exchange
$J$, and EDSR tone amplitudes $\Omega_1$ and $\Omega_2$. Bottom: Accompanying phase modulation of the EDSR tones.}\label{fig:VDSsingeshotshape}
\end{figure}

\begin{figure}[t!]
  \centering
  \includegraphics[width=.9\columnwidth]{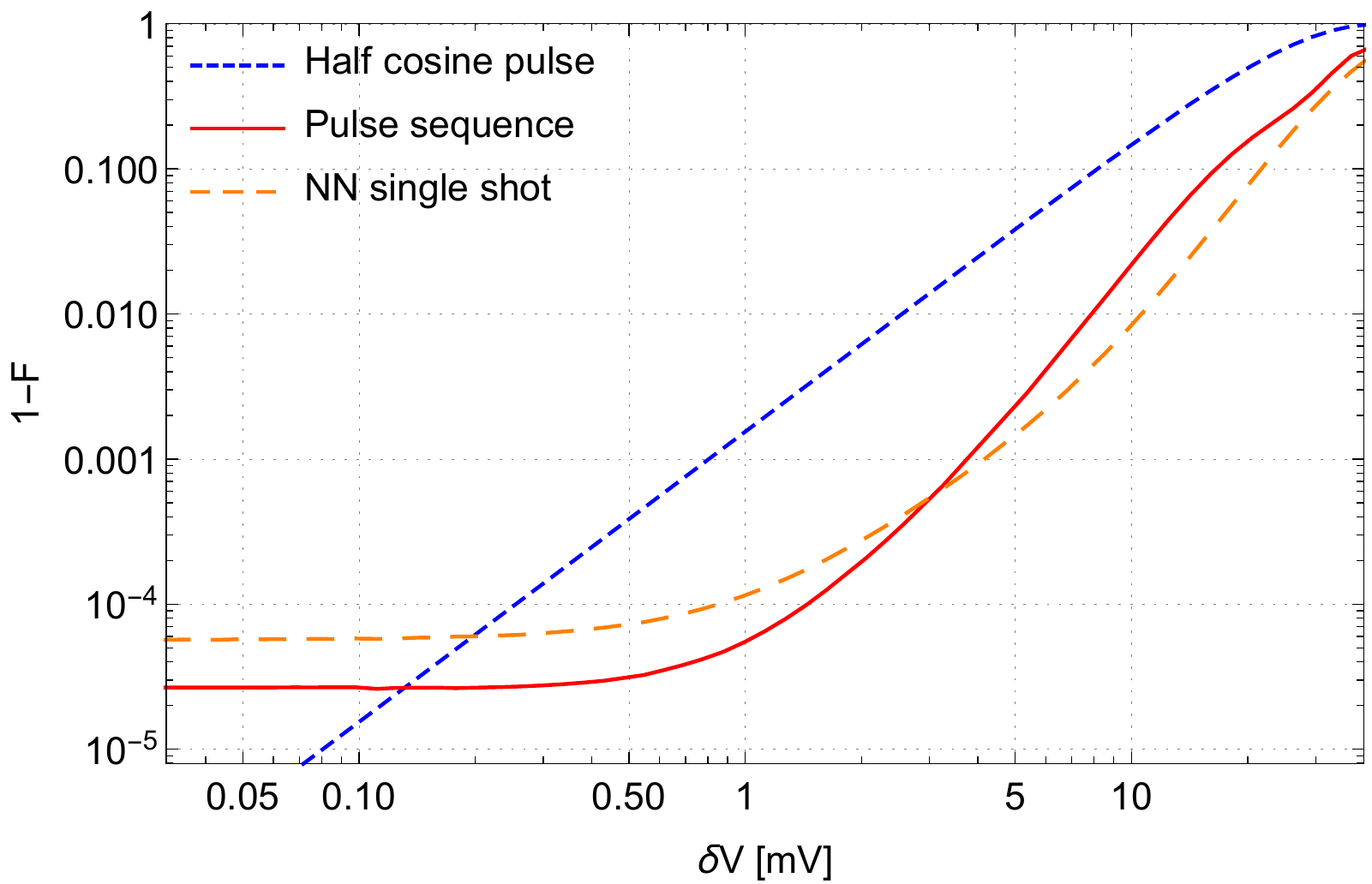}\\
  \includegraphics[width=.9\columnwidth]{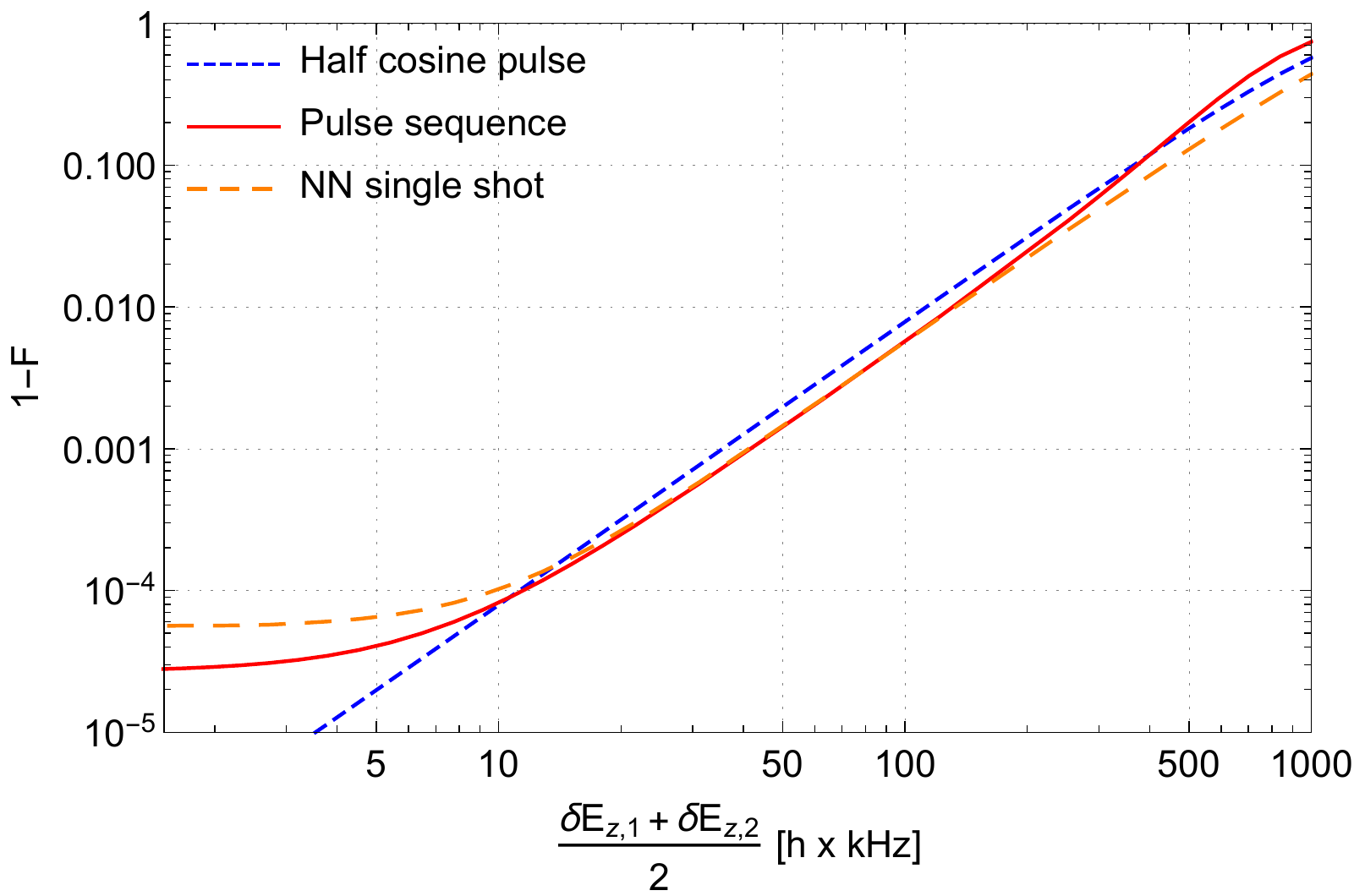}
  \caption{Top: Infidelity vs barrier fluctuations, $\delta V$. Bottom: Infidelity vs average Zeeman fluctuation.}\label{fig:VDSfidelity}
\end{figure}

With such a large maximum Rabi frequency and maximum exchange it is very fast to implement the pulse sequence from Ref.~\cite{utkan2018}. To this end the EDSR amplitude is ramped as $(1-\cos(\pi t/t_r))/2$ up to its maximum amplitude with $t_r=10$ns as shown in Fig.~\ref{fig:VDSpulseshape}. The exchange $J(V)$ is also shown in Fig.~\ref{fig:VDSpulseshape} and is pulsed adiabatically during the $i$th segment with the form:
\begin{equation}
    J_i(t)=J_{\text{max},i} \frac{\tanh(\frac{t}{T_{\text{Ramp},i}}) \tanh(\frac{T_{\text{Ad},i} - t}{T_{\text{Ramp},i}})}{\tanh^2(\frac{T_{\text{Ad},i}}{2 T_{\text{Ramp},i}})}
\end{equation}
where $t$ is the time as measure from the beginning of the segment, $J_{\text{max},i}$ sets the height of the peak, $T_{\text{Ad},i}$ is the total duration of the pulse segment, and $T_{\text{Ramp},i}$ is the ramp time. $J_{\text{max},i}$ was manually chosen such that an optimization over the ramp time yielded the lowest total time while still having an adiabatic infidelity, as defined in Sec.~\ref{subsec:nn}, below $10^{-6}$. For the $e^{i \frac{\eta}{2}ZZ}$ pulse section $J_{\text{max},1}=h \times 4.5$ MHz, $T_{\text{Ramp},1}=0.095\mu$s and $T_{\text{Ad},1}\approx0.109\mu$s. For the $e^{i \frac{\pi}{4}ZZ}$ pulse section $J_{\text{max},2}=h \times 7$ MHz, $T_{\text{Ramp},1}\approx0.12\mu$s and $T_{\text{Ad},1}\approx0.107\mu$s.
The resulting pulse sequence is $\approx 1.12 \mu s$ long which is about about 11 times longer than the uncorrected half cosine pulse. To create a shorter robust pulse the neural network was used which resulted in a single shot pulse which is $0.5 \mu s$ long. The shape of the control fields for the single shot pulse are represented in Fig.~\ref{fig:VDSsingeshotshape}. Although the single shot pulse has the benefit of being faster than the pulse sequence by a factor of about 2, the exchange pulse is non-adiabatic and requires more bandwidth. The single-shot pulse has a 3dB Bandwidth of approximately $28$MHz which is limited by the ESR Bandwidth. Additionally, when applying a hard cutoff low-pass filter, a cutoff frequency of $770$MHz was required for an infidelity of $10^{-3}$ to be maintained in the absence of noise. The 3dB bandwidth is well below the bandwidth modern arbitrary waveform generators such as the $640$MHz at 1.6GS/s samplerate accessible with the Tektronix 5014C used in Ref.~\cite{AWG}. Although a hard cutoff low-pass filter required more than $640$MHz it is not such a large difference that such a value might be capable of being reached in the future.
The resulting infidelity versus quasistatic noise for both pulses is represented in Fig.~\ref{fig:VDSfidelity}. The infidelity of a half cosine-shaped pulse for an uncorrected {\sc cz}-gate as used in Ref.~\cite{xue2021} is also plotted for comparison. For the uncorrected gate to have an infidelity of $10^{-4}$ requires fluctuations of barrier gate voltage to be less than $0.3$mV, whereas the pulse sequence or single shot pulse bring that threshold up to $1.5$mV or $1$mV, respectively. 

\begin{figure}[t!]
    \centering
    \includegraphics[width=.9\columnwidth]{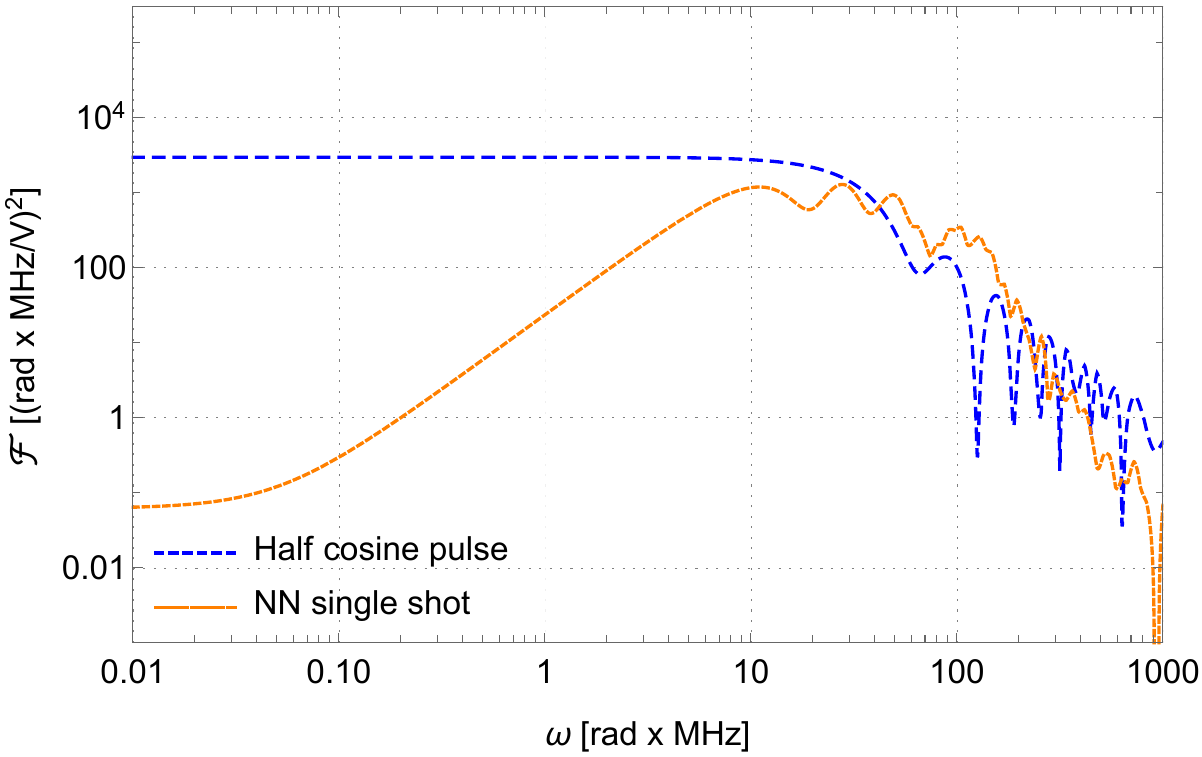}
    \includegraphics[width=.9\columnwidth]{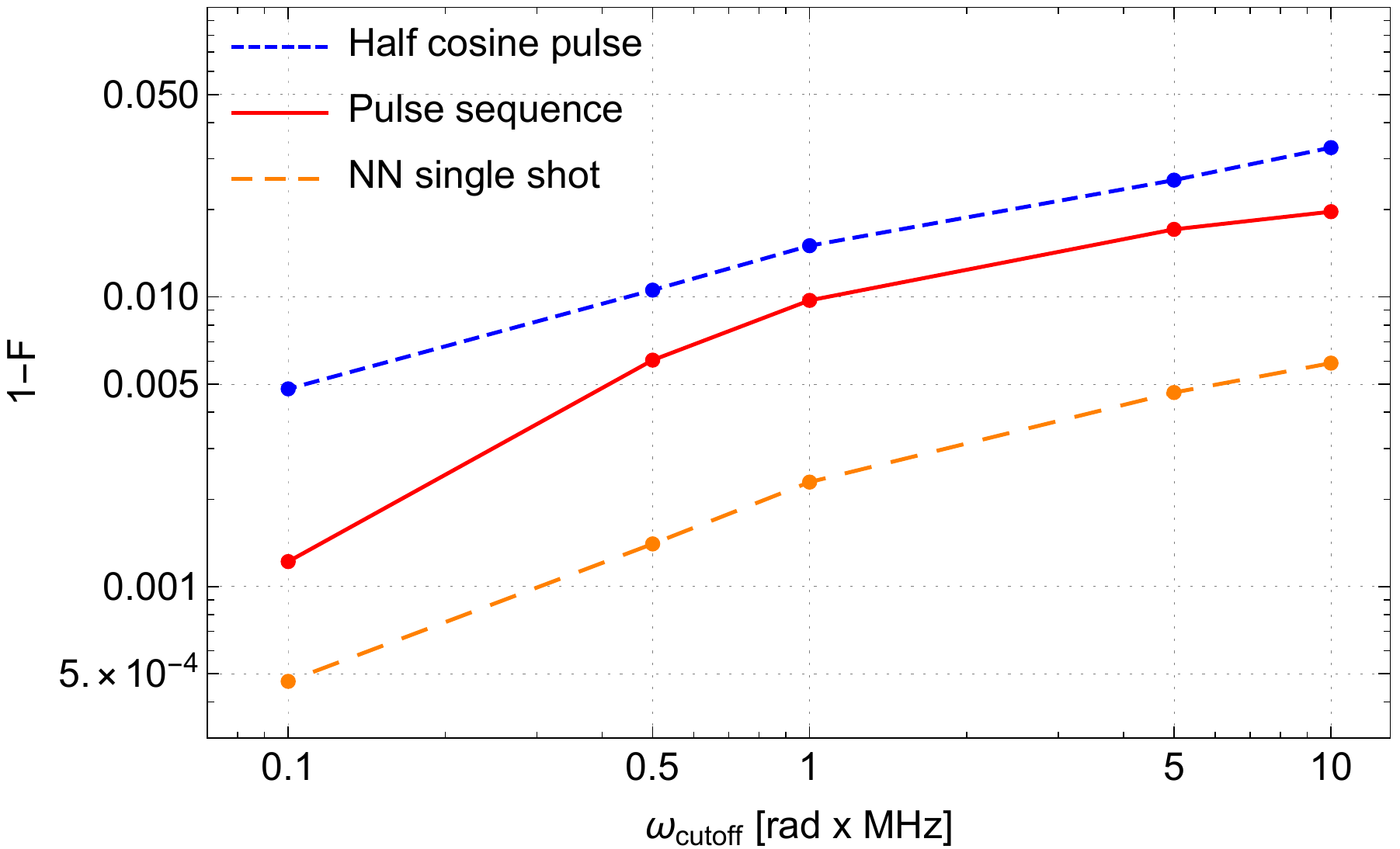}
    \caption{Top: Filter function, $\mathcal{F}(\omega)$, for the charge noise,  of the single shot pulse and the half cosine pulse versus noise frequency $\omega$ in the top plot. Bottom: Average fidelity of the pulse sequence, the single shot pulse and the half cosine pulse vs cutoff frequency, with daily calibration and $\delta V_{rms}=3$mV.}
    \label{fig:VDSfilter}
\end{figure}
All three gate methods are similarly sensitive to Zeeman energy fluctuations, with an infidelity under $10^{-4}$ for fluctuations below $\sim h \times 13$kHz. The sequence and single shot pulses are not designed to be robust against fluctuations in Zeeman energy, as the sequence's single-qubit gates are not robust against these fluctuations and the single shot pulse was not optimized for these fluctuations.

These results however do not take the typical $1/f$ frequency dependence of charge noise into account. To analyze how non-static noise affects the fidelity, we calculate the filter function, $\mathcal{F}(\omega)$, using the method from Ref.~\cite{Green_2013} with appropriate multiplicative noise weights $\chi$ for fluctuations in $V$ from the formalism in Ref.~\cite{utkan2018}. The filter function is plotted in Fig.~\ref{fig:VDSfilter} for the half cosine pulse and the single shot pulse. We have not included results for the pulse sequence on Fig.~\ref{fig:VDSfilter} because the perturbative filter function method requires the error in the Hamiltonian be linearized in $\delta V$, which is not possible at $V=0$ (which is the case for most of the pulse sequence) for errors of the form $(V+\delta V)^{\gamma}$, so the method is not directly applicable in that situation. 

To compare all three gate methods, the infidelity as a result of frequency-dependent noise is calculated by averaging over 500 simulations of $1/f$ noise realization for $\delta V$ with a particular infrared cutoff frequency, $\omega_{\text{ir}}$, and high-frequency cutoff, $\omega_{\text{cutoff}}$, of the noise. $1/f$ noise realizations were generated using a sum of random telegraph processes with switching rates logarithmically distributed from $\omega_{\text{ir}}$ to $\omega_{\text{cutoff}}$. This results in a $1/f$ noise spectrum between $\omega_{\text{ir}}$ and $\omega_{\text{cutoff}}$, with a flat spectrum below that range and a $1/f^2$ spectrum above it \cite{Yang2016}. The resulting noisy signal is then shifted and have a mean of 0 over the calibration time and rescaled to have a standard deviation of $\delta V_{rms}=3$mV, which is the noise strength where the single shot pulse and the pulse sequence have the same quasistatic infidelity of $5 \times 10^{-3}$ in Fig.~\ref{fig:VDSfidelity}. The average fidelity of the three pulses is plotted in Fig.~\ref{fig:VDSfilter} as a function of $\omega_{\text{cutoff}}$ with $\omega_{\text{ir}}=10^{-5} \text{rad} \times $Hz (i.e., daily calibration).
This shows that for $1/f$ noise in this frequency band (which extends up to the largest frequency in the control Hamiltonian), the pulse sequence as well as the single shot pulse improve the infidelity. However, the robustness of gates produced with the neural network approach relies on the specific shape of the functions $J(V)$ and $f_i(V)$. For example, for the single-shot pulse if $\gamma$ is decreased the infidelity in the absence of error scales quadratically with the decrease resulting in a less robust pulse. At a 2\% decrease in $\gamma$ the infidelity of the single-shot pulse in the absence of error becomes $6.2\times 10^{-3}$. This means that an accurate model of the system is required.

Another experimental concern could be the needed precision of the EDSR amplitude to achieve high fidelity. In the absence of noise, an infidelity of $10^{-3}$ is still achieved with relative errors in EDSR amplitude of 0.3\% for the single shot pulse and 0.4\% for the pulse sequence, a practical level for typical arbitrary waveform generators \cite{vanDijk2019}.

\subsection{ESR device with non-negligible crosstalk}
\label{subsec:ESR}
Now we consider a two-qubit device with non-negligible crosstalk. We present two approaches. First, we consider again using the analytical composite pulse sequence of Eq.~\eqref{eq:seqfull}, but this time with each individual segment optimized using a deep neural network as discussed in Sec.~\ref{subsec:nn}. Second, we try dropping the pulse sequence structure and creating a single-shot pulse entirely via the neural network. 

The Zeeman splitting at the barrier gate idle voltage is taken to be $\Delta E_z= h \times 7.05$MHz \cite{UNSWprivate} and is comparable in size to the typical values of exchange (a few MHz) and ESR Rabi frequency (nearly a MHz), so it is not possible to apply the RWA. Consequently the full rotating frame Hamiltonian must be used. For two-tone resonant driving the Hamiltonian for this system has the form
\begin{widetext}
\begin{multline}\label{eq:Hdriven}
H(t)= \frac{J(t)}{4} \left[ZZ + \cos \Delta E_z t \left(XX+YY\right) + \sin \Delta E_z t \left(XY-YX\right) \right]
+ V\left(\frac{d E_{z,1}}{d V}ZI + \frac{d E_{z,2}}{d V}IZ \right)
\\
+ \frac{\Omega_1(t)}{2}\left[ \cos\phi_1(t) XI +\sin\phi_1(t) YI +  \frac{1}{r_{12}} \cos \left(\frac{\Delta E_z t}{2}+\phi_1(t)\right) IX + \frac{1}{r_{12}}\sin \left(\frac{\Delta E_z t}{2}+\phi_1(t)\right) IY\right]
\\
+ \frac{\Omega_2(t)}{2}\left[\cos\phi_2(t) IX + \sin\phi_2(t) IY + r_{12} \cos \left(-\frac{\Delta E_z t}{2}+\phi_2(t)\right) XI +r_{12} \sin \left(-\frac{\Delta E_z t}{2}+\phi_2(t)\right) YI \right]
\end{multline}
\end{widetext}
where the nonuniform strength of the ESR signal across the sample results in a ratio of $r_{12}=0.812$ between the driving of the two dots and the barrier gate voltage adjustment functions are $f_i(V)=V \frac{d E_{z,i}}{d V} $ \cite{veldhorst_Dzurak}. The exchange depends of the barrier gate voltage as $J(V)=J_\text{off} e^{\frac{V-V_\text{off}}{V_0}}$ where $J_\text{off}=h \times 0.928$kHz, $V_\text{off}=-0.101$V and $V_0=0.0266$V\cite{UNSWprivate}. The maximum ESR strength for the modeled device is $\Omega_\text{max}=507$kHz for the first qubit and $\Omega_\text{max}=625$kHz for the second qubit because of the nonuniform ESR signal strength. The values of the derivatives in the barrier gate adjustment functions are $\frac{d E_{z,1}}{d V} =h \times 7.07$MHz/V and $\frac{d E_{z,2}}{d V}=h \times 5.05$MHz/V. The shape of $\Omega_i$ and $\phi_i$ are determined with the neural network to result in the necessary robust pulses.

\subsubsection{Optimized segments in a composite pulse}\label{subsec:segments}
First, we present shaped exchange pulses so as to obtain a shortcut to adiabaticity during the entangling segments of Eq.~\eqref{eq:seqfull}. The speed up for these pulses is about $1.5$ times faster than pulses with the same adiabatic infidelity, as defined in Sec.~\ref{subsec:nn}, found by limiting the flip probability determined by perturbation theory as in Ref.~\cite{utkan2018}. However the improvement in the total pulse time is not very large and similar performance can be found if the slower adiabatic pulses, which require less bandwidth, are used. The ESR line is not driven during these segments.
The adiabatic shape of $J(V)$ is determined by a neural network with a modified cost function. The robustness term $\mathcal{E}$ in the cost function for the adiabatic $J$ was replaced with the adiabatic infidelity multiplied by a weighting factor of 0.1.

\begin{figure}
  \centering
  \includegraphics[width=.9\columnwidth]{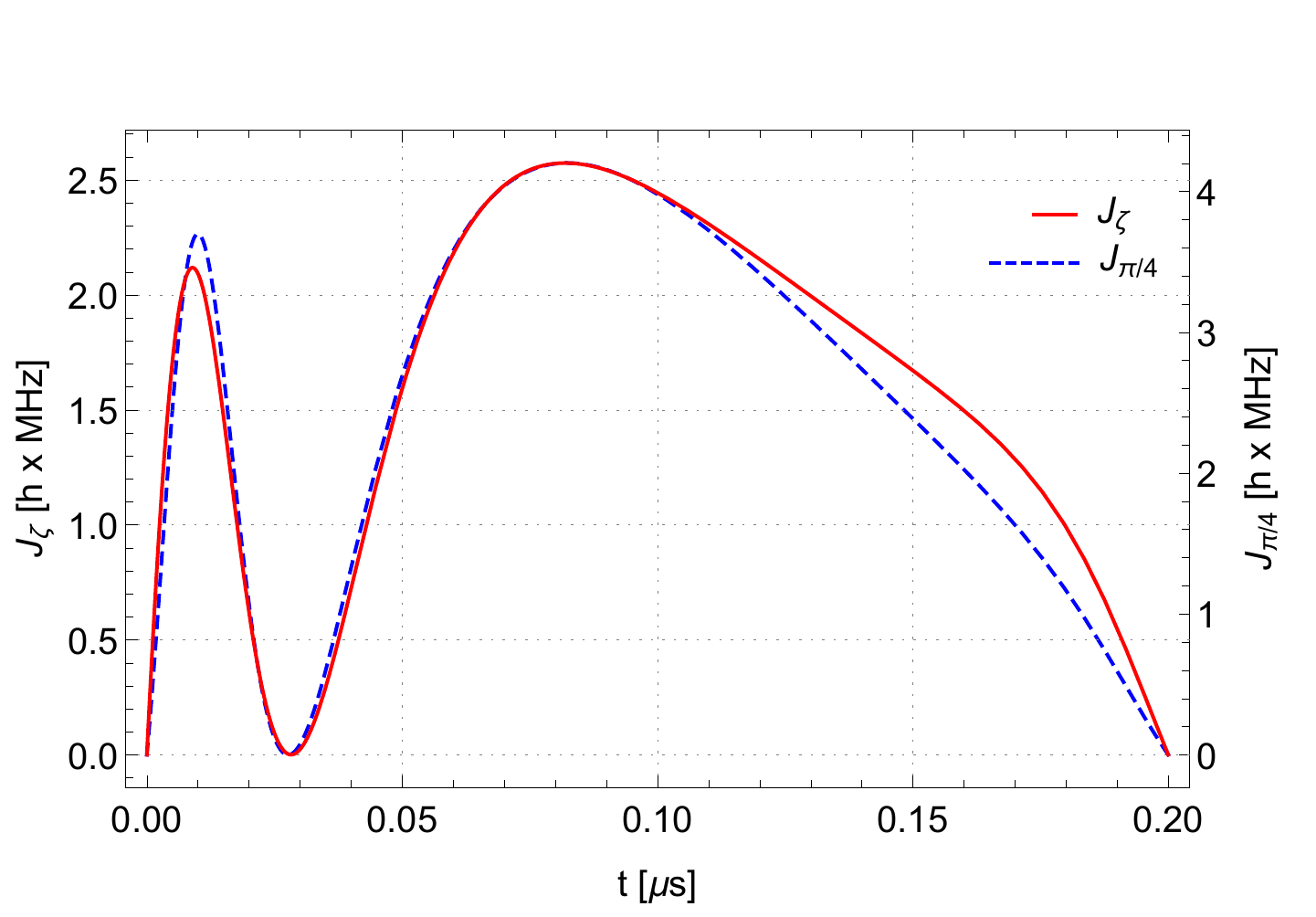}
  \caption{Shaped exchange pulses $J_{\pi/4}(t)$ and $J_{\zeta}(t)$ for an adiabatic $e^{-i \frac{\pi}{4} ZZ}$ and rotation represented by the dashed blue line and the solid red line. The strength of the exchange $J_{\zeta}(t)$ on the left side of the plot is scaled by a factor of $\approx 1.63$ compared to the strength of the exchange $J_{\pi/4}(t)$ shown on the right.}\label{fig:czshape}
\end{figure}

The $e^{- i \frac{\pi}{4} ZZ}$ segment of Eq.~\eqref{eq:seqfull} is performed by pulsing $J$ as $J_{\pi/4}(t)$ shown in Fig.~\ref{fig:czshape}. All neural networks for pulse shapes and pulse segments determined by neural networks as well as  the accompanying single-qubit rotations (cf.~Eq.~\eqref{eq:costtarget}), $\varphi_i$, are included in the Supplemental Material \cite{supplement}.
Similarly, the $e^{- i \frac{\zeta}{2} ZZ}$ segment is performed by pulsing $J$ as $J_{\zeta}(t)$ also shown in Fig.~\ref{fig:czshape} with its vertical axis rescaled by $\approx 1.63$. The two pulse shapes in Fig.~\ref{fig:czshape} are not simply related by a scale factor because the $f_i(V)$ functions on terms in Eq.~\eqref{eq:Hdriven}.  

Now we present shaped ESR pulses to perform the single-qubit rotation segments of Eq.~\eqref{eq:seqfull} such that they are robust against charge noise, which, along with the structure of the composite sequence itself, renders the overall entangling gate robust. The exchange coupling is turned off during these segments while two-tone ESR driving is used. 

\begin{figure}
  \centering
  \includegraphics[width=.9\columnwidth]{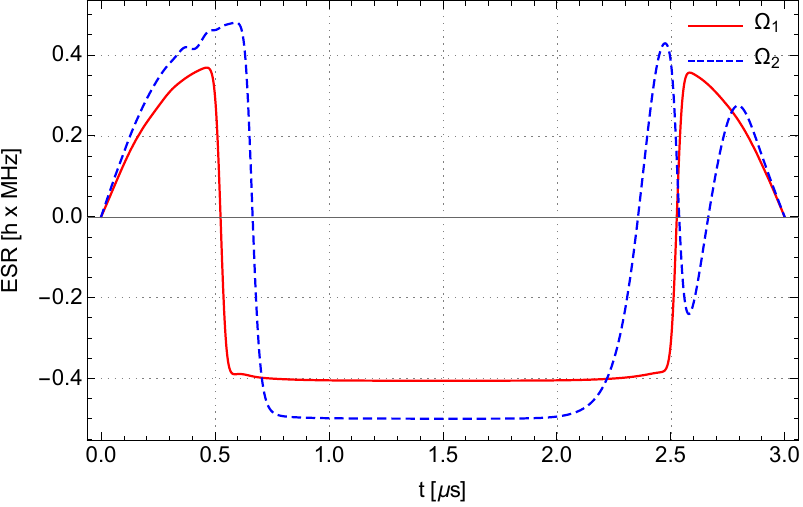}
  \includegraphics[width=.9\columnwidth]{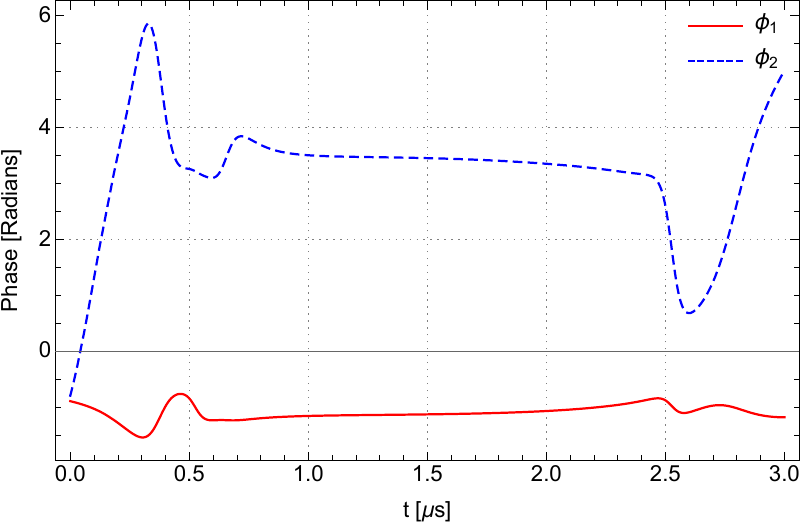}
  \caption{Shaped pulse for an $XX$ gate robust against Zeeman fluctuations. Top: Two-tone ESR amplitude modulation. Bottom: Accompanying phase modulation.}\label{fig:XXshape}
\end{figure}

\begin{figure}
  \centering
  \includegraphics[width=.9\columnwidth]{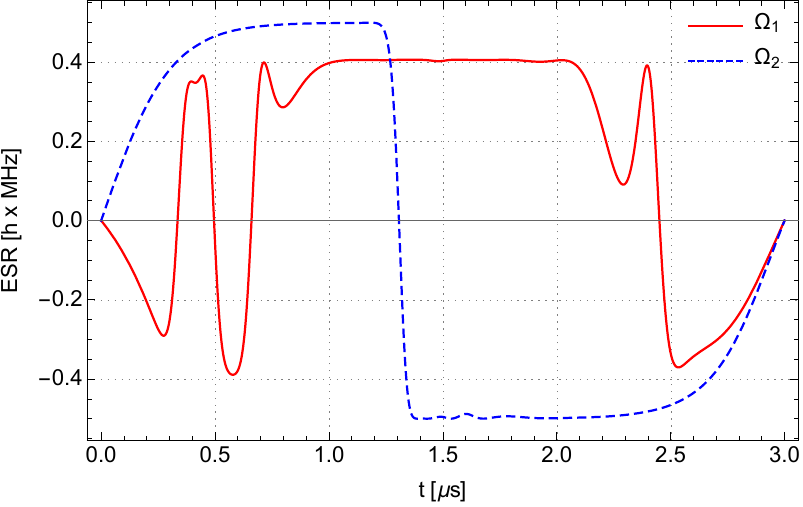}
  \includegraphics[width=.9\columnwidth]{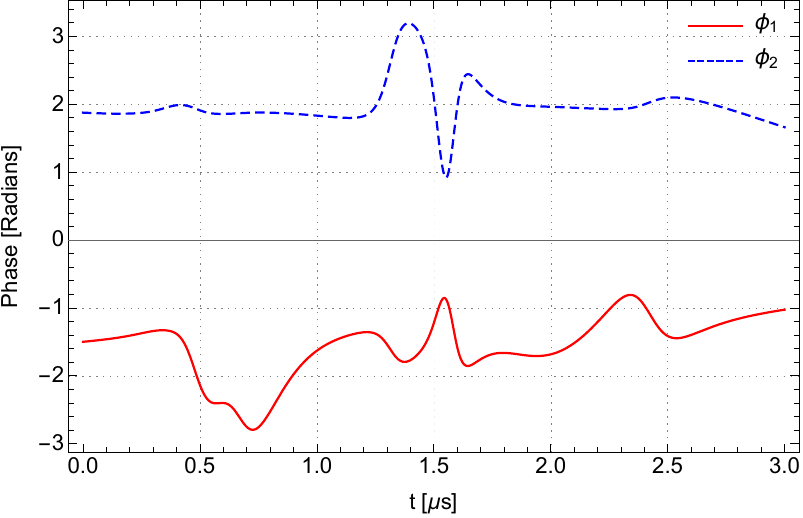}
  \caption{Shaped pulse for a $e^{-i \left(\frac{\pi}{2}XI + \frac{ \theta-\pi}{2} IX\right)}$ rotation robust against Zeeman fluctuations. Top: Two-tone ESR amplitude modulation. Bottom: Accompanying phase modulation.}\label{fig:theta1shape}
\end{figure}

The robust $XX$ segment is shown in Fig.~\ref{fig:XXshape}, where the amplitude and phase of each tone is simultaneously modulated. The $XX e^{-i \left(\frac{ \theta}{2} IX\right)}$ segment is performed in one pulse and is shown in Fig.~\ref{fig:theta1shape}. Additionally, $XX e^{i \frac{ \theta}{2} IX}=e^{i\frac{\pi}{2}IZ}(XX e^{-i \frac{ \theta}{2} IX})e^{i\frac{\pi}{2}IZ}$, so that segment of the sequence can also be performed by pulsing as in Fig.~\ref{fig:theta1shape} and accounting for the additional $z$ rotations in software. 

\begin{figure}
  \centering
  \includegraphics[width=.9\columnwidth]{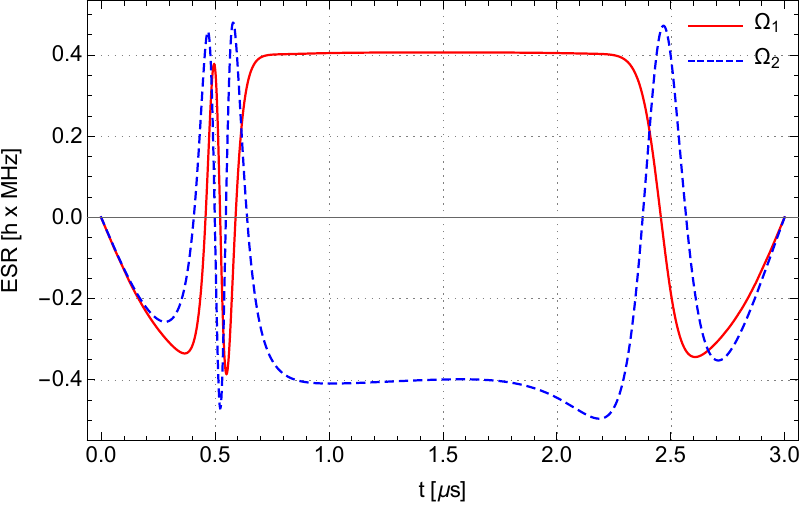}
  \includegraphics[width=.9\columnwidth]{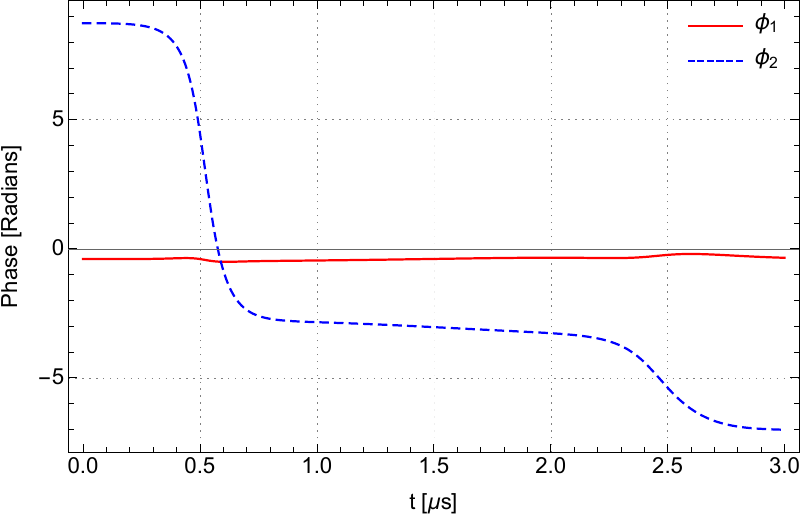}
  \caption{Shaped pulse for a $e^{-i \left(\frac{\pi}{2}XI - \frac{\pi+ \eta}{2} IX\right)}$ rotation robust against Zeeman fluctuations. Top: Two-tone ESR amplitude modulation. Bottom: Accompanying phase modulation.}\label{fig:pietashape}
\end{figure}

\begin{figure}
  \centering
  \includegraphics[width=.9\columnwidth]{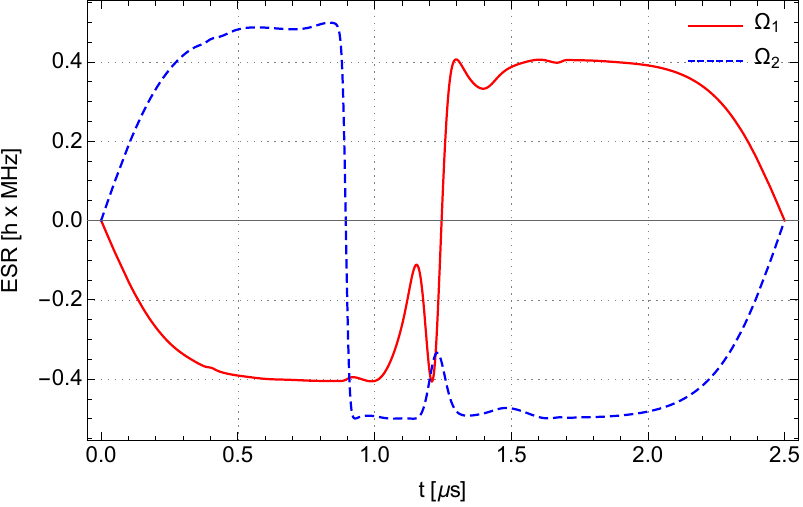}
  \includegraphics[width=.9\columnwidth]{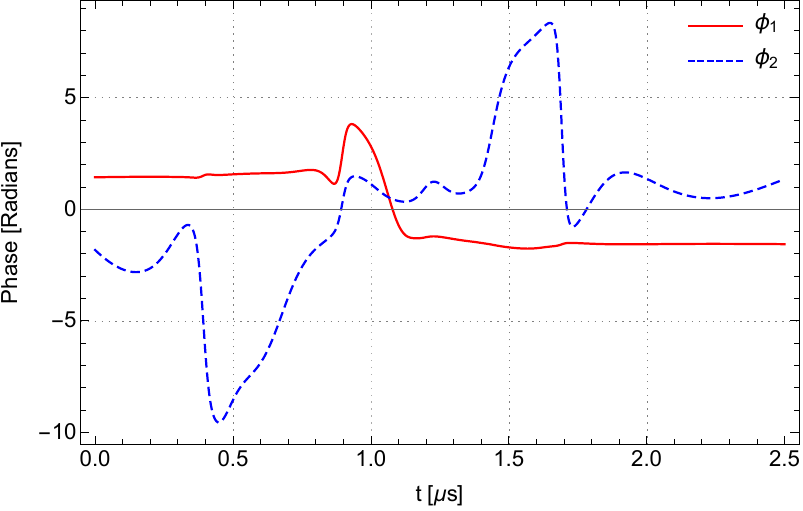}
  \caption{Shaped pulse for a $e^{-i \frac{\eta}{2} IX}$ rotation robust against Zeeman fluctuations. Top: Two-tone ESR amplitude modulation. Bottom: Accompanying phase modulation.}\label{fig:etashape}
\end{figure}

The $XX e^{i \left(\frac{ \eta}{2} IX\right)} $ segment is likewise performed in one pulse and shown in Fig.~\ref{fig:pietashape}. Finally the robust $e^{i \frac{\eta}{2} IX}$ pulse is performed as shown in Fig.~\ref{fig:etashape}.

\begin{figure}
  \centering
  \includegraphics[width=.9\columnwidth]{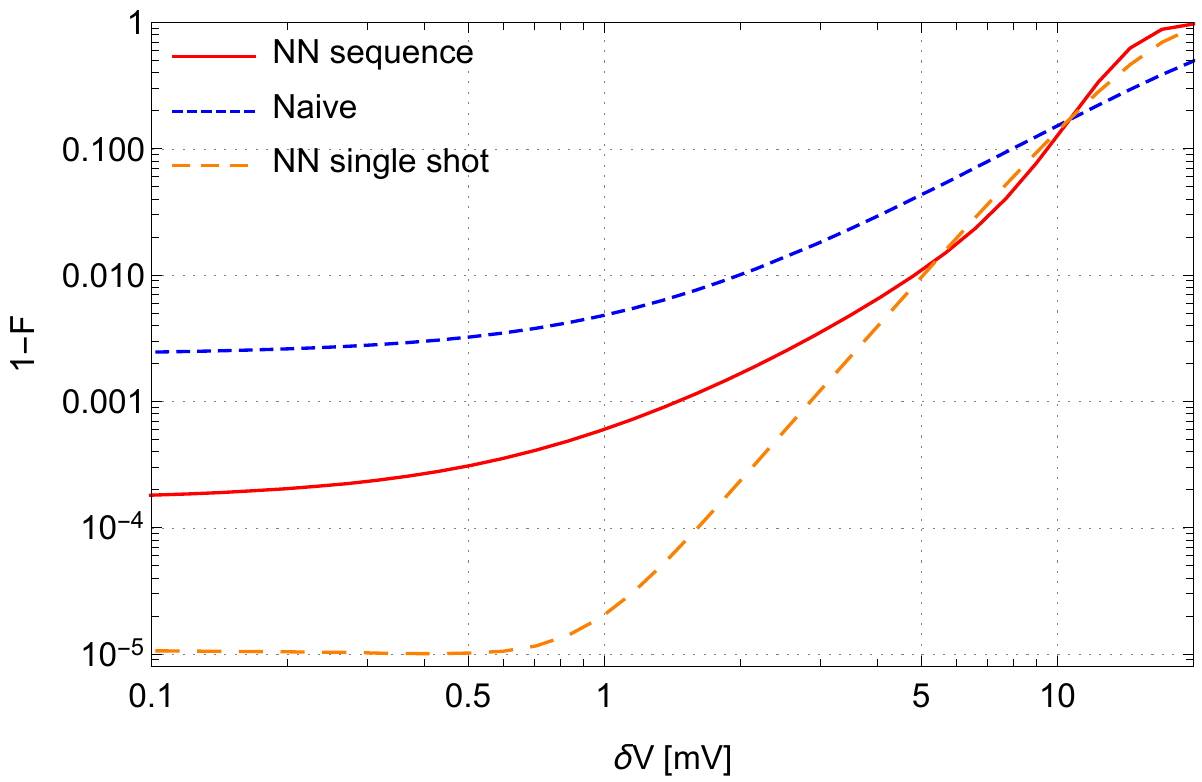}
  \includegraphics[width=.9\columnwidth]{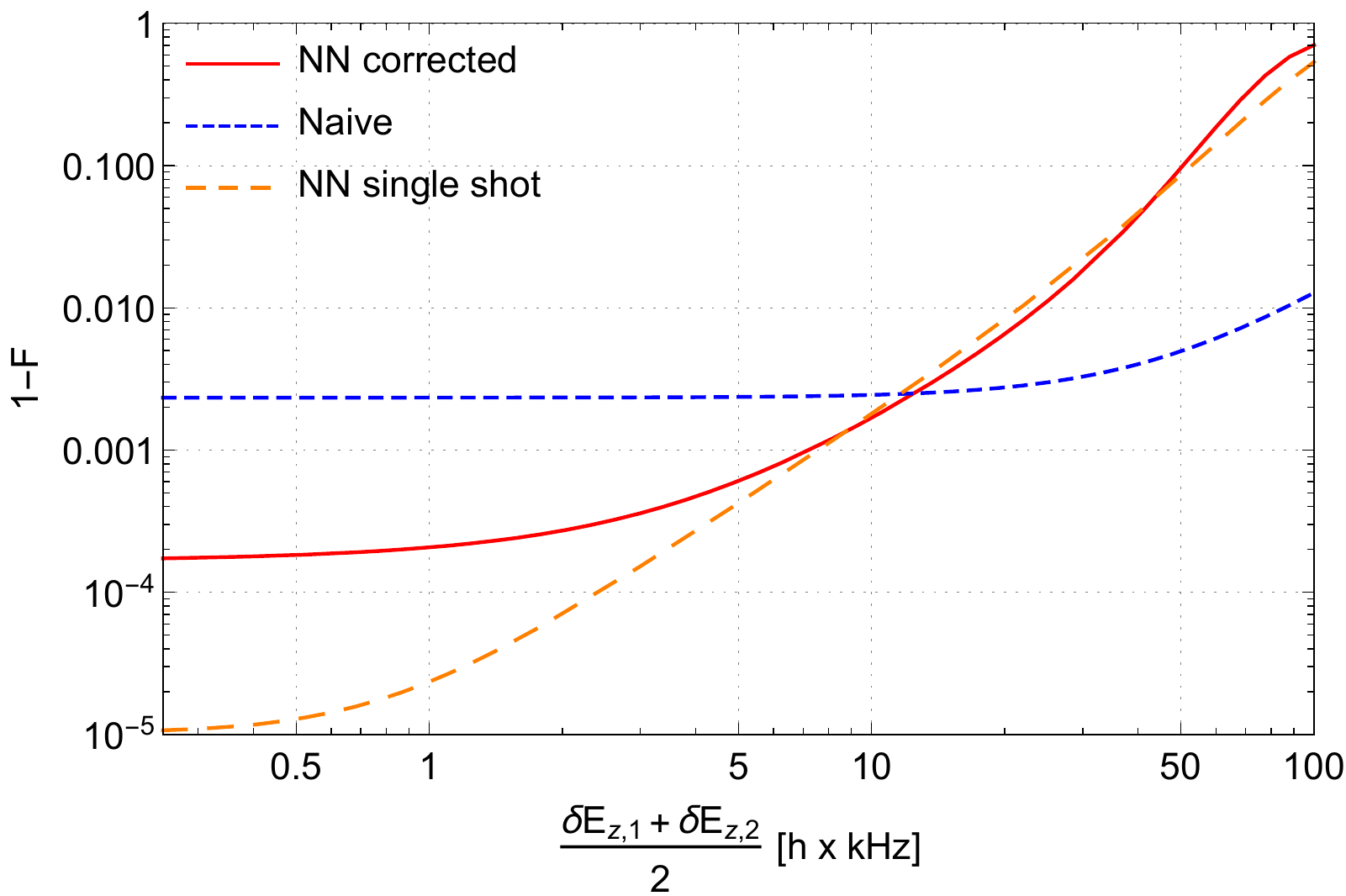}
  \caption{Infidelity of the neural network optimized pulse sequence vs quasistatic fluctuations in barrier gate voltage (top) and in average Zeeman energy (bottom). For comparison the naive square pulse infidelity is also plotted.}\label{fig:seqfid}
\end{figure}

The infidelity of the entire pulse sequence versus barrier gate voltage fluctuations and Zeeman fluctuations is shown in Fig.~\ref{fig:seqfid}. (The ``single shot" pulse is discussed in the next subsection.) For comparison, Fig.~\ref{fig:seqfid} also includes the infidelity of a naive square pulse with exchange strength $J=\frac{\Delta E_z}{\sqrt{3}}$ and duration $T=\frac{3}{2 \Delta E_z}$ to create a {\sc cz} gate. An infidelity below $10^{-3}$ is achieved  for fluctuations in $V$ below $1.2$mV or fluctuations in average Zeeman energies of both qubits below $h \times 6$kHz. This is an improvement over the naive square pulse which never achieves an infidelity of $10^{-3}$ due to the crosstalk.
The entire pulse sequence requires $1.2\mu$s of entangling time and about $20.5\mu$s of single-qubit rotation time for a total sequence time of $21.7\mu$s. One could shave off $5.5\mu$s by omitting the outer $x$-rotations, which do not change the entangling properties of the gate, at the cost of not having a standard {\sc cz} gate. The infidelity is also limited by the time-dependence of the noise since charge noise is typically $1/f$ noise and not quasistatic as assumed above. In this section the effect of $1/f$ noise will be estimated instead of directly simulated as in Sec.~\ref{subsec:EDSR} because the pulses (especially the pulse sequence) are longer and more complicated, and are thus numerically more difficult to simulate with rapid fluctuations.
It is easy to estimate a limit on the infidelity of $1-\exp(-(T_\text{gate}/T_{2,\text{Hahn}})^2) \sim 3\times 10^{-4}$ \cite{Cywinski2008}, for $T_{2,\text{Hahn}}=1.2$ms \cite{Veldhorst_2014}. This could be improved greatly by device improvements such as increasing the maximum ESR drive strength as well as a longer $T_{2,\text{Hahn}}$. In particular the infidelity would decrease as the inverse square of the Rabi frequency which is the main bottleneck for this approach.

\subsubsection{Simultaneous single-shot shaped pulses on $J$ and ESR}
It is desirable to find a faster robust entangling gate than the 22$\mu$s composite sequence of the previous subsection. One might think that an option would be to just use the shorter sequence of Eq.~\eqref{eq:seq} along with faster, nonrobust single-qubit rotations. That would indeed reduce the gate time down to about 4$\mu$s, but, although the shorter sequence is robust against fluctuations in exchange coupling, the $g$-factor fluctuations due to charge noise ruin the overall robustness of the sequence.

\begin{figure}
  \centering
  \includegraphics[width=.9\columnwidth]{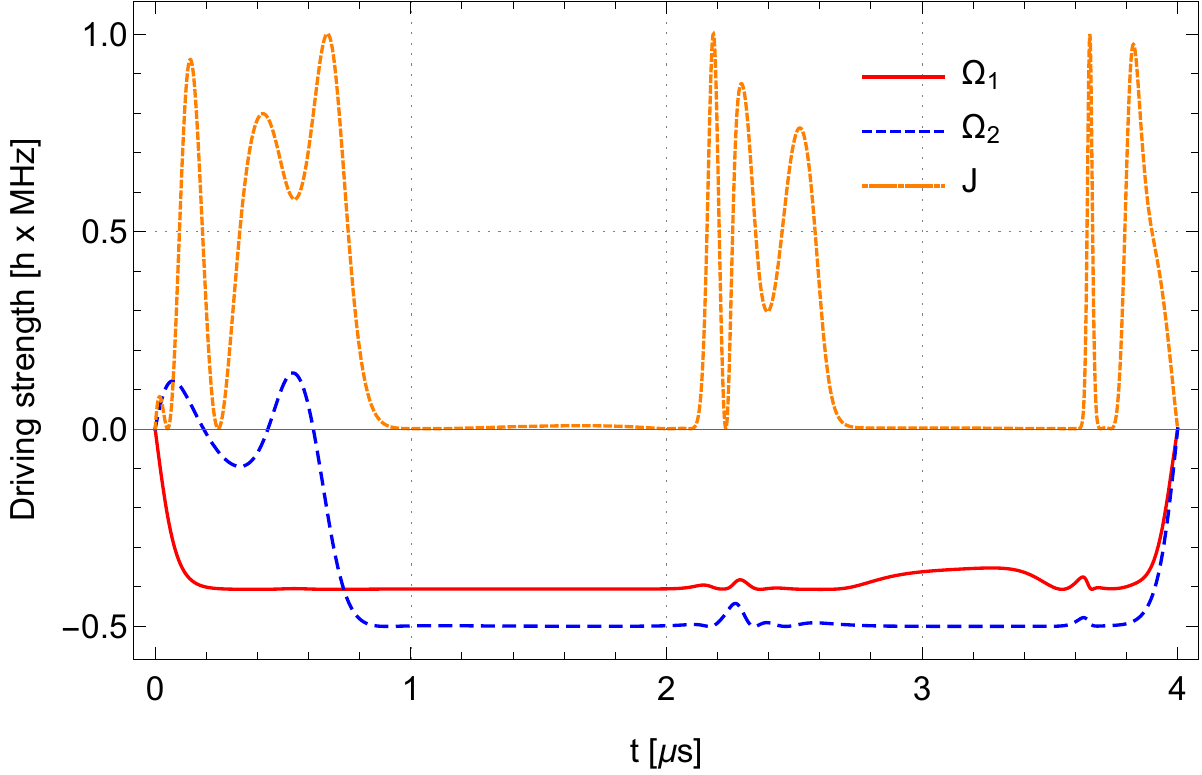}
  \includegraphics[width=.9\columnwidth]{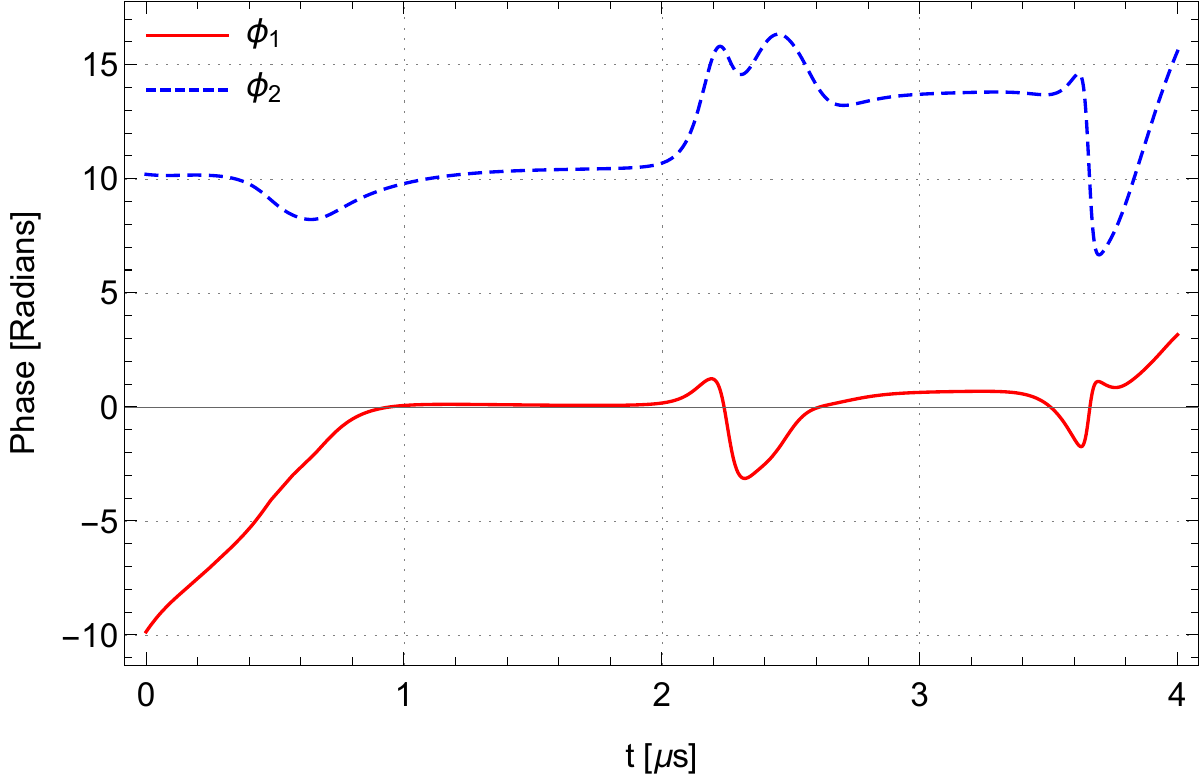}
  \caption{Simultaneous shaped pulse of exchange and two-tone ESR. Top: Amplitudes vs time. Bottom: Accompanying phase modulation of the ESR tones.}\label{fig:fullpulseshape}
\end{figure}

In this section we demonstrate an alternative to the prior approach of numerically optimizing the individual segments of a composite sequence. We show that the desired {\sc cz} gate can be performed faster by abandoning the analytically derived composite pulse sequence structure of alternating ESR-only and $J$-only pulses, instead directly carrying out the more demanding computational task of optimizing simultaneous pulse shapes on both. The evolution operator was calculated using Eq.~\ref{eq:Hdriven} and the synchronized pulse shapes obtained via neural network optimization are shown in Fig.~\ref{fig:fullpulseshape}.
The pulse is shorter than the sequence by about a factor of 5 as it is only $4\mu s$ long. The improvement in performance can be seen in the fidelity plots shown in Fig.~\ref{fig:seqfid} where the infidelity scales better with low Zeeman energy fluctuations than the pulse sequence and is more robust against barrier fluctuations. An infidelity of $10^{-4}$ is achieved below $1.6$mV fluctuations in barrier voltage and below $h \times 5$kHz Zeeman energy fluctuations. The quicker nature of this pulse results could lead to better performance depending on the decoherence time. A simple estimation using $T_{2,\text{Hahn}}=1.2$ms \cite{Veldhorst_2014} limits the infidelity to $1-F\approx 1-\exp(-(T/T_{2,\text{Hahn}})^2)\approx10^{-5}$ \cite{Cywinski2008}. The downside of this pulse is that it requires more bandwidth for $J$ since it is no longer driven adiabatically. The single-shot pulse has a 3dB Bandwidth of approximately $18.4$MHz which is limited by the ESR Bandwidth. Additionally, when applying a hard cutoff low-pass filter, a cutoff frequency of $170$MHz was required for an infidelity of $10^{-3}$ to be maintained in the absence of noise. This is well below the bandwidth available in modern arbitrary waveform generators such as the $640$MHz at 1.6GS/s samplerate accessible with the Tektronix 5014C used in Ref.~\cite{AWG},
Another experimental concern could be the needed precision of the ESR amplitude to achieve high fidelity. In the absence of noise an infidelity of $10^{-3}$ is still achieved in with an error in ESR amplitude of 0.9\% for the single shot pulse and 0.5\% for the pulse sequence, which are again experimentally realistic tolerances  \cite{vanDijk2019}.
Although the pulse is robust to charge noise in the presence of crosstalk as defined in the introduction, other types of crosstalk, such as heating and capacitive interactions, are not taken into account because the mechanisms that cause most these effects are not well-characterized and thus are not incorporated in our model. However, if the results of these effects can be quantified in the model Hamiltonian (like the barrier-dependent resonance shift, $f_i(V)$, included in this work), neural network optimization could be reapplied in the presence of these effects. The exchange floor $J_0$ can cause unwanted $ZZ$ rotation while doing an uncorrected $X$-gate which can be called crosstalk. However, the neural network optimization takes $J_0$ into account in the Hamiltonian and can correct this type of crosstalk as well. 

\section{Conclusion}
In this paper we have shown that it is possible to use a neural network to optimize a robust pulse or pulse sequence to generate {\sc cz} gates robust against crosstalk even in the presence of noise. 

For the case of small crosstalk and large maximum EDSR driving (Sec.~\ref{subsec:EDSR}), our approach yielded infidelities of $10^{-4}$ for quasistatic fluctuations in $V$ below $1.2$mV or Zeeman energy fluctuations below $h \times 12$kHz. While this is a large improvement over a naive square pulse, it takes significantly longer at over $1.2\mu$s. To circumvent the lengthy pulse sequence we showed a $0.5\mu$s single shot pulse optimized directly with the neural network which still showed large improvement over the naive pulse.

For the case of large crosstalk and low maximum ESR strength (Sec.~\ref{subsec:ESR}) it was shown that it is possible to optimize the pulse sequence from Ref.~\cite{utkan2018}, but it was more efficient to optimize the whole problem with the neural network since it was about 5 times faster. The resulting pulse is limited in fidelity by the ratio of the gate time, determined by the ESR amplitude $\Omega_\text{max}$, to the timescale on which the charge noise switches, $T_2$. For predominantly low-frequency noise, our approach gives order-of-magnitude improvements in the infidelity due to charge noise.

\section*{Acknowledgements} \label{sec:acknowledgements}
    This research was sponsored by the Army Research Office (ARO), and was accomplished under Grant Number W911NF-17-1-0287.

\bibliographystyle{apsrev4-1} 
\bibliography{refs} 
\end{document}